# Heat treatment effects on the superconducting properties of Ag-doped SrKFeAs compound

ZHANG ZhiYu, WANG Lei, QI YanPeng, GAO ZhaoShun, WANG DongLiang, ZHANG XianPing, MAYanWei∗

Key Laboratory of Applied Superconductivity, Institute of Electrical Engineering, Chinese Academy of Sciences, P. O. Box 2703, Beijing 100190, China

The superconducting properties of polycrystalline $Sr_{0.6}K_{0.4}Fe_2As_2$ were strongly influenced by Ag doping (*Supercond. Sci. Technol. 23 (2010) 025027*). Ag addition is mainly dominated by silver diffusing, so the annealing process is one of the essential factors to achieve high quality Ag doped $Sr_{0.6}K_{0.4}Fe_2As_2$. In this paper, the optimal annealing conditions were studied for Ag doped $Sr_{0.6}K_{0.4}Fe_2As_2$ bulks prepared by a one-step solid reaction method. It is found that the annealing temperature has a strong influence on the superconducting properties, especially on the critical current density $J_c$. As a result, higher heat treatment temperature (~900$^o$C) is helpful in diffusing Ag and reducing the impurity phase gathered together to improve the grain connectivity. In contrast, low-temperature sintering is counterproductive for Ag doped samples. These results clearly suggest that annealing at ~900$^o$C is necessary for obtaining high $J_c$ Ag-doped samples.

---

∗ Author to whom correspondence should be addressed; E-mail: ywma@mail.iee.ac.cn

## 1. Introduction

The recently discovered quaternary arsenide oxide superconductor La(O$_{1-x}$F$_x$)FeAs (1111 phase) with the superconducting critical transition temperature ($T_c$) of 26 K [1], has stimulated an enormous interest in the field of superconductivity. Soon after, the superconducting transition temperature ($T_c$) was increased to 55 K by replacing La with Sm[2]. Nearly at the same time, Oxygen-free FeAs-based compounds with the ThCr$_2$Si$_2$-type (122 phase) structure exhibit superconductivity with $T_c$ as high as 38 K, such as (Ba$_{1-x}$K$_x$)Fe$_2$As$_2$, (Sr$_{1-x}$K$_x$)Fe$_2$As$_2$[3,4]. The 1111 family, indeed, shows higher $T_c$, but it also needs a higher sintering temperature (nearly 1200$^o$C), while the 122 family is less anisotropic and exhibits very high upper critical field as well as very low processing temperature required[5-7]. So these 122 type iron-based pnictide superconductors have attracted much attention.

As is shown in the other superconductors, silver has been widely used in the fabrication of superconductors and also as dopants or additives to improve the microstructure and superconducting properties. Previously our group studied the effect of silver addition on the polycrystalline Sr$_{0.6}$K$_{0.4}$Fe$_2$As$_2$ samples and found that the critical current density $J_c$ and the irreversibility field were obviously increased[8]. More recently, Wang *et al.* reported that significant transport critical currents had been achieved first in Sr$_{0.6}$K$_{0.4}$Fe$_2$As$_2$ wires and tapes when silver was used as addition and sheath material[9]. Unlike the cases of hole-doping or electron-doping, silver serves as fillers filling the gaps of micro-cracks or pores in the microstructure and improving connectivity between grains[8]. However, to our knowledge, so far rare systematic reports are found on the relationship between sample preparation techniques, microstructure and superconducting properties of polycrystalline pnictide superconductors. Actually the critical properties of superconductors are strongly influenced by heat treatment conditions. The performance of $YB_2C_3O_y$ and $MgB_2$ superconductors can be further enhanced by optimizing the heat treatment temperatures[10,11]. Therefore, it is necessary to investigate the heat treatment effect on the Ag doped $Sr_{0.6}K_{0.4}Fe_2As_2$ superconductor.

In this work, the effects of heat-treatment temperatures, varied from 700 to $900^oC$, on the microstructure, $J_c$, $H_{irr}$ and $H_{c2}$ of 20 wt% Ag doped $Sr_{0.6}K_{0.4}Fe_2As_2$ are investigated.

## 2.  Experimental details

The polycrystalline pure and 20 wt% Ag doped $Sr_{0.6}K_{0.4}Fe_2As_2$ bulk samples were prepared by a one-step PIT method developed by our group[12]. High-purity Sr filings, Fe powder, As and K pieces are used as starting materials. The raw materials were ground by ball milling in the Ar atmosphere for more than 15 hours. In order to compensate the loss of elements during the milling and sintering procedures, extra 10 % As and 10 % K were added. After the ball milling, the raw powders were added with 20 wt% Ag powder (200 mesh, purity 99.9%). Then, the powders were ground in a mortar for 15 minutes. The final powders were encased and sealed into Nb tubes. Finally, the samples were heat-treated at 700–$900^oC$ for 35hrs. The argon gas was allowed to flow into the furnace during the heat-treatment process to reduce the oxidation of the samples.

The crystal structure and phase purity of all samples were treated by an X-ray diffraction (XRD) using Cu－Kα radiation from 10 to 80°. The superconducting properties of the wires were studied by dc magnetization and four-probe resistivity measurements using a physical property measurement system (Quantum Design PPMS) in magnetic fields up to 9 T. The magnetic critical current density $J_c$ was determined using the extended Bean model[13], with the formula $J_c$ (= $20\Delta m/Va(1-a/3b)$) (for estimating the global $J_c$ of bulk materials), where $\Delta m$ is the width of the magnetization (M) loop; $a$ and $b$ are the length and width of the bulk ($a < b$), and $V$ is the volume of sample. Rectangular specimens nearly in the same dimension of $3\times2\times1$ mm$^3$ were cut from the samples. The field was applied normal to the long axis (3mm) of the sample. The microstructure and composition of the sample were analyzed using a scanning electron microscope (SEM) equipped with an energy-dispersive x-ray (EDX) analysis.

## 3. Results and discussion

The X-ray diffraction patterns of all samples prepared at different temperatures are presented in Figure 1. As can be seen, almost all main peaks for all the samples can be indexed on the basis of tetragonal ThCr$_2$Si$_2$-type structure with the space group I4/mmm. However, some Ag, and a small amount of FeAs and AgSrAs impurity phases were observed in all the samples. It can be noted that the content of the second phases is decreased by increasing the sintering temperature, especially for AgSrAs, which nearly disappeared in the samples sintered at 900°C. On the other hand, the relative intensity of the Ag peaks has no significant change with the increase of heat treatment temperature.

Figure 2 shows the temperature dependence of resistivity at the zero magnetic field for samples heated at different temperatures. Clearly, for all the samples the resistance drops at 35 K and the zero resistance reaches 33 K. This result indicates that the annealing temperatures ranging from 700 to 900$^o$C have no effect on the $T_c$ of Ag doped $Sr_{0.6}K_{0.4}Fe_2As_2$ samples.

Figures 3 (a) and (b) show magnetic $J_c$ derived from the magnetic hysteresis for various samples at 5 and 20 K. As can be seen, for Ag doped samples, the $J_c$ monotonically increased with the increase of the heat treatment temperature, and got the top value for the samples sintered at 900$^o$C. A notable feature of the $J_c$–$B$ behavior is the peak effect presented in the samples sintered at 850 and 900$^o$C in Figure 3 (b), and a similar fishtail effect has been observed in the single crystal $Ba_{0.6}K_{0.4}Fe_2As_2$ reported by Yang $et$ $al.$[7] and 1111-type iron-based superconductors reported by Zhao $et$ $al.$[14]. In general, this peak effect is helpful in improving the flux pinning behavior of superconductors in high-fields.

Figure 3 (c) shows the field dependence of magnetic $J_c$ at 20 K and 6 T on the sintering temperature, for pure and Ag doped $Sr_{0.6}K_{0.4}Fe_2As_2$ samples. As we can see, the value of $J_c$ increases gradually with the increase of the annealing temperature. Both pure and Ag doped $Sr_{0.6}K_{0.4}Fe_2As_2$ have the best $J_c$ for a high sintering temperature, with the highest $J_c$ value of around 2400A/cm$^2$ obtained for Ag doped $Sr_{0.6}K_{0.4}Fe_2As_2$ sintered at 900°C. However, it should be noted that the high $J_c$ value of pure samples almost saturates around 850$^o$C. In contrast, for the Ag doped samples, the $J_c$ increases rapidly from 850 to 900°C, and the $J_c$ value of Ag doped samples finally exceeds the pure samples at 900°C. The result reveals that the effect of Ag on the $J_c$ property of $Sr_{0.6}K_{0.4}Fe_2As_2$ is strongly correlated to the heat treatment temperature. The function of silver is to raise the $J_c$ value of $Sr_{0.6}K_{0.4}Fe_2As_2$ when the heat treatment temperature is ~900°C. However, low-temperature sintering is counterproductive for the Ag doped samples. A similar variation was obtained for all the other fields between 0 and 7 T, and also at 5 K. So the information obtained from the present experiment shows that the heat treatment temperature is a major agent affecting silver in $Sr_{0.6}K_{0.4}Fe_2As_2$ samples.

Figure 4 shows the resultant *H-T* phase diagrams for the Ag-doped samples sintered at 800 and 900°C. We carried out the resistance versus temperature measurements on the Ag–added bulk samples in various magnetic fields by the four-probe resistive method. The onset temperature decreases very slowly with the increasing magnetic field. However, the zero-resistance temperature drops quickly to low temperatures in both samples (not shown). The 10% and 90% points on the resistive transition curves were used to define the $H_{irr}$ and $H_{c2}$. The result shows that heat treatment temperature has little effect on $H_{c2}$. A striking feature of the $H_{c2}(T)$ curve is the rapid increase in $H_{c2}$ with a decreasing temperature, suggesting a very high upper critical field near T= 0 K. The slope of $H_{c2}$ near $T_c$ (99% Rn from R-T), $dH_{c2}/dT|T_c$, is ~6.3T/K. The corresponding $H_{c2}(T=0)$ value derived from the Werthamer-Helfand-Hohenberg formula is-$0.693T_c(dH_{c2}/dT ) = 150T$[15]. In contrast to the $H_{c2}(T)$ behavior, the irreversibility field $H_{irr}$ shows a significant increase with the increase of heat treatment temperature. In particular, the $H_{irr}$ value for the Ag-doped samples sintered at 900°C is 5.8 T at 32 K, much larger than 0.6 T for the Ag-added samples sintered at 800°C.

A further study on the microstructure of the Ag-doped samples was performed using SEM. Figures 5 (a)-(e) show the SEM images of the Ag doped $Sr_{0.6}K_{0.4}Fe_2As_2$ samples sintered at 700-900°C. It is observed that the grain size obviously increases as the sintering temperature rises. At the same time, many flocculent substances can be clearly seen in samples sintered at a lower temperature and gradually disappear with the increase of heat treatment temperature.

In order to get more information about the microstructure of Ag-doped samples, we did EDX mapping experiments, the SEM image and elemental maps of samples sintered at 700 and 900°C were shown in Fig.6 and Fig.7, respectively. All elements were detected, which indicates that the major phase is $Sr_{0.6}K_{0.4}Fe_2As_2$. It can be clearly seen the distribution of silver is inhomogeneous in the whole area when the samples are sintered at a low sinter temperature (700°C). At the same time, the impurity phase AgSrAs is formed together due to the gathered silver, which seriously limits the connectivity of grain. When the heat temperature increases, the silver spread more evenly and the impurity phases disappear which contribute to the improvement of connectivity. Our group previously reported that the glassy phase as well as the amorphous layer presented in almost all the grain edges and boundaries in $Sr_{0.6}K_{0.4}Fe_2As_2$ samples can be suppressed by Ag addition[8]. These lead to the concluded that the silver could improve the connectivity of $Sr_{0.6}K_{0.4}Fe_2As_2$ samples when the heat treatment temperature is ~900$^{\rm o}$C.

## 4. Conclusions

We investigated the effects of the heat treatment temperature on the phase transformation, $T_c$, $J_c$, $H_{c2}$ and $H_{irr}$ for 20 wt% Ag doped polycrystalline $Sr_{0.6}K_{0.4}Fe_2As_2$ bulks. It is observed that $H_{irr}$ and $J_c$ increased with increasing heat treatment temperature. However the heat treatment temperature had no influence on $T_c$ and $H_{c2}$.

When compared to the pure $Sr_{0.6}K_{0.4}Fe_2As_2$ as a reference sample, it was found that the $J_c$ value of Ag doped sample was much larger than the pure sample if the heat treatment temperature was ~900$^{\rm o}$C, but the $J_c$ was deteriorated by Ag doping when the sinter temperature was lower than 900$^{\rm o}$C. On the other hand, the Ag addition enhanced the $J_c$ of samples at a higher sinter temperature (~900$^{\rm o}$C), resulting from the diffusion of silver and the disappearance of gathered impurity phase. The data indicate that the silver can improve the superconducting properties of $Sr_{0.6}K_{0.4}Fe_2As_2$ samples when the heat treatment temperature is 900$^{\rm o}$C.

## Acknowledgements

The authors thank Profs. Haihu Wen, Liye Xiao and Liangzhen Lin for their help and useful discussions. This work was partially supported by the Beijing Municipal Science and Technology Commission (Grant No. Z09010300820907). The National Science Foundation of China (Grant No. 50802093) and the National '973' Program (Grant No. 2006CB601004).

**Captions**

Figure 1 X-ray diffraction patterns of the Ag-doped $Sr_{0.6}K_{0.4}Fe_2As_2$ samples processed at different annealing temperatures.

Figure 2 Superconducting transition temperature dependence of resistivity for the Ag doped samples at different annealing temperatures.

Figures 3 (a) and (b) Magnetic field dependene of $J_c$ at 5 K and 20K for the Ag doped samples at different annealing temperatures. (c) Dependence of $J_c$ at 20 K and 6 T on the sintering temperature for pure and Ag doped $Sr_{0.6}K_{0.4}Fe_2As_2$ samples.

Figure 4 The upper critical field line $H_{c2}$ and $H_{irr}$ as a function of the temperature for Ag doped $Sr_{0.6}K_{0.4}Fe_2As_2$ samples sintered at 800 and 900$^o$C.

Figure 5 High magnification SEM micrographs for the Ag-doped samples at different annealing temperatures. (a)700$^o$C (b)750$^o$C (c)800$^o$C (d)850$^o$C (e)900$^o$C

Figure 6 EDX mapping images of the Ag-doped $Sr_{0.6}K_{0.4}Fe_2As_2$ samples sintered at 700$^o$C.

Figure 7 EDX mapping images of the Ag doped $Sr_{0.6}K_{0.4}Fe_2As_2$ samples sintered at 900$^o$C.

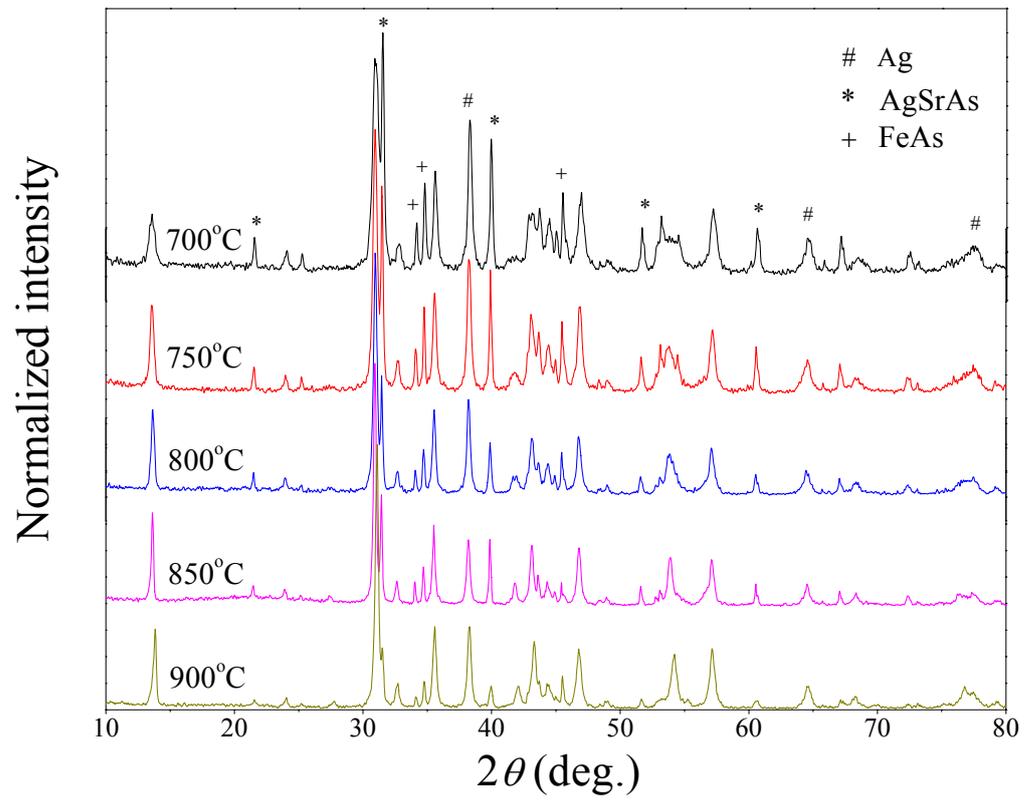

Fig.1 Zhang et al.

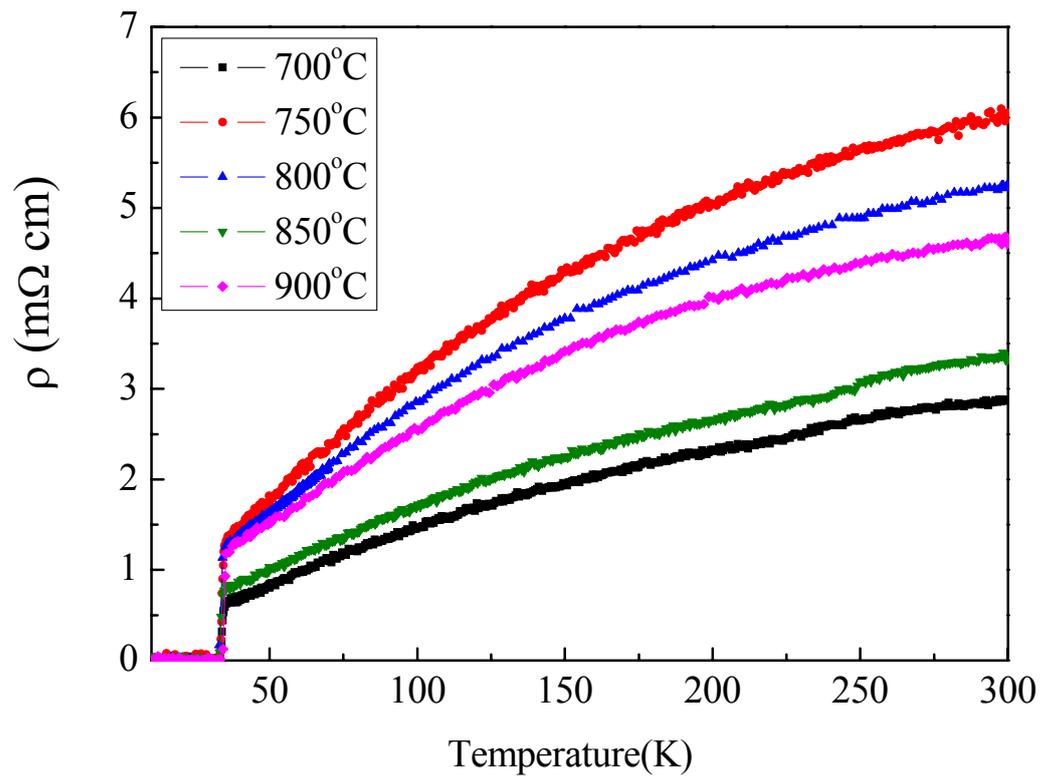

Fig.2 Zhang et al.

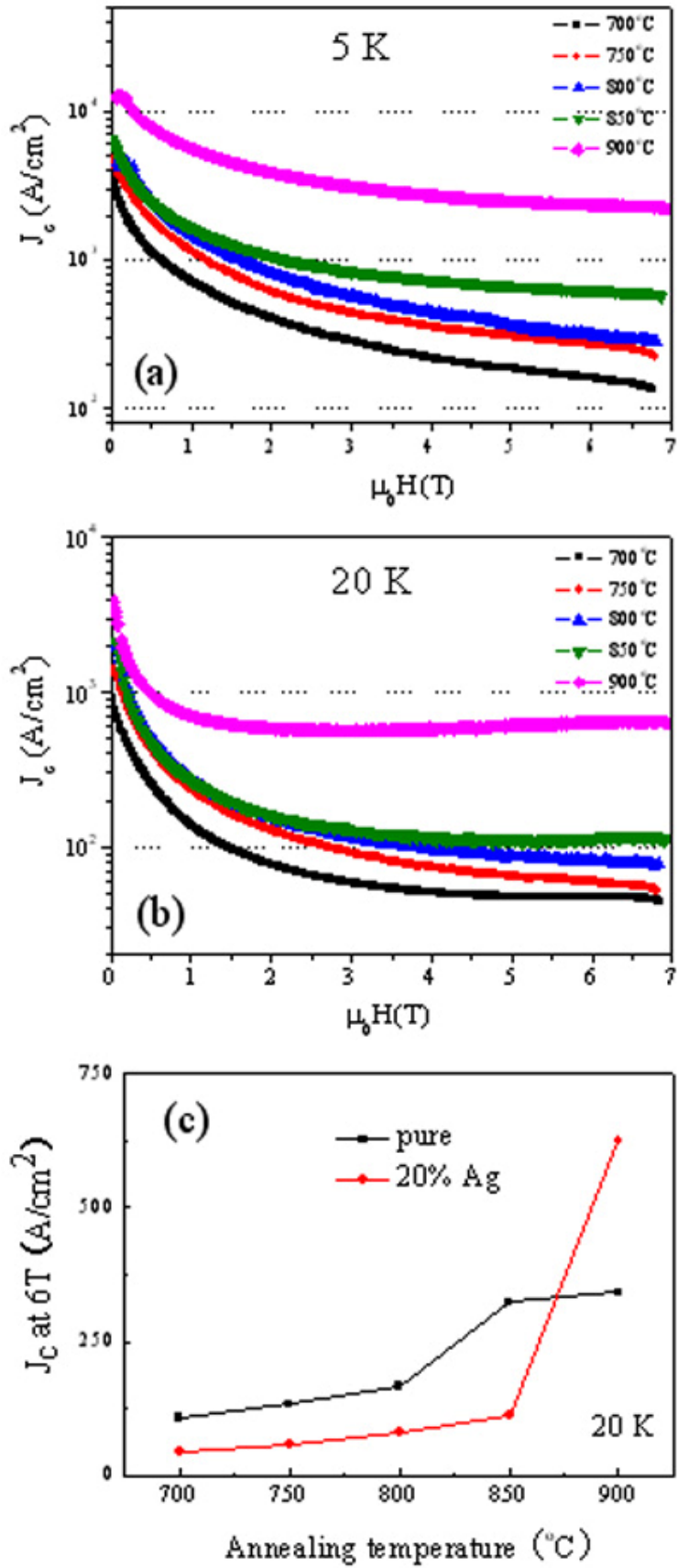

Fig.3 Zhang et al.

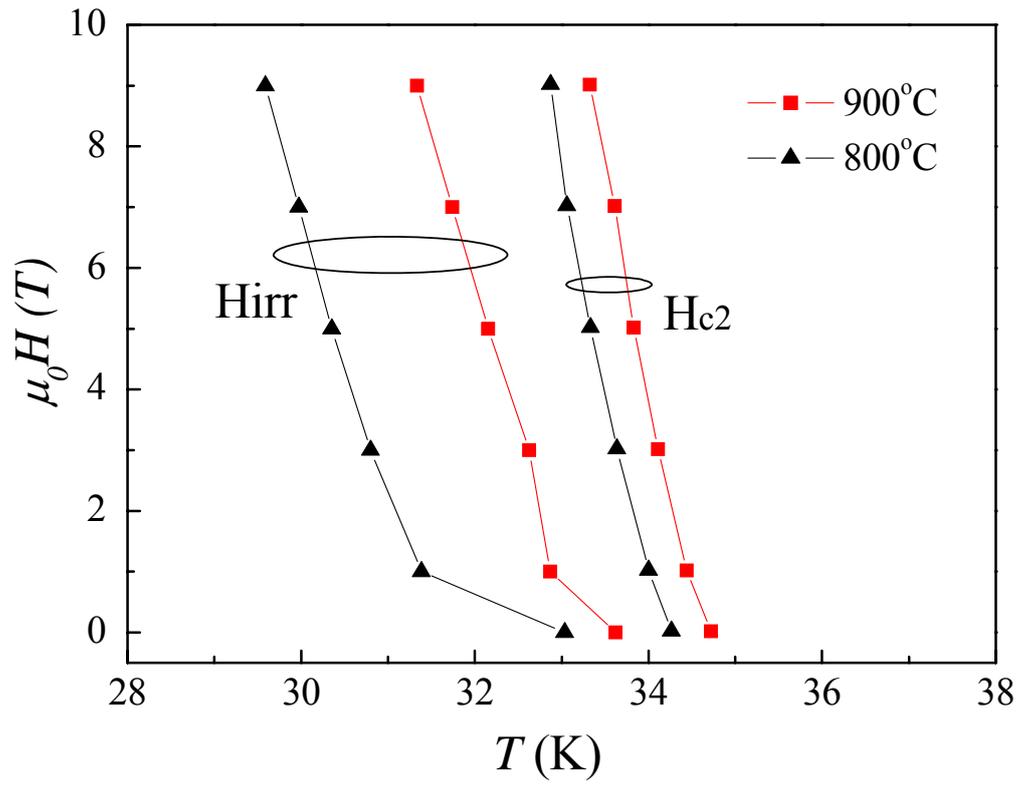

Fig. 4 Zhang et al.

Fig.5 Zhang et al.

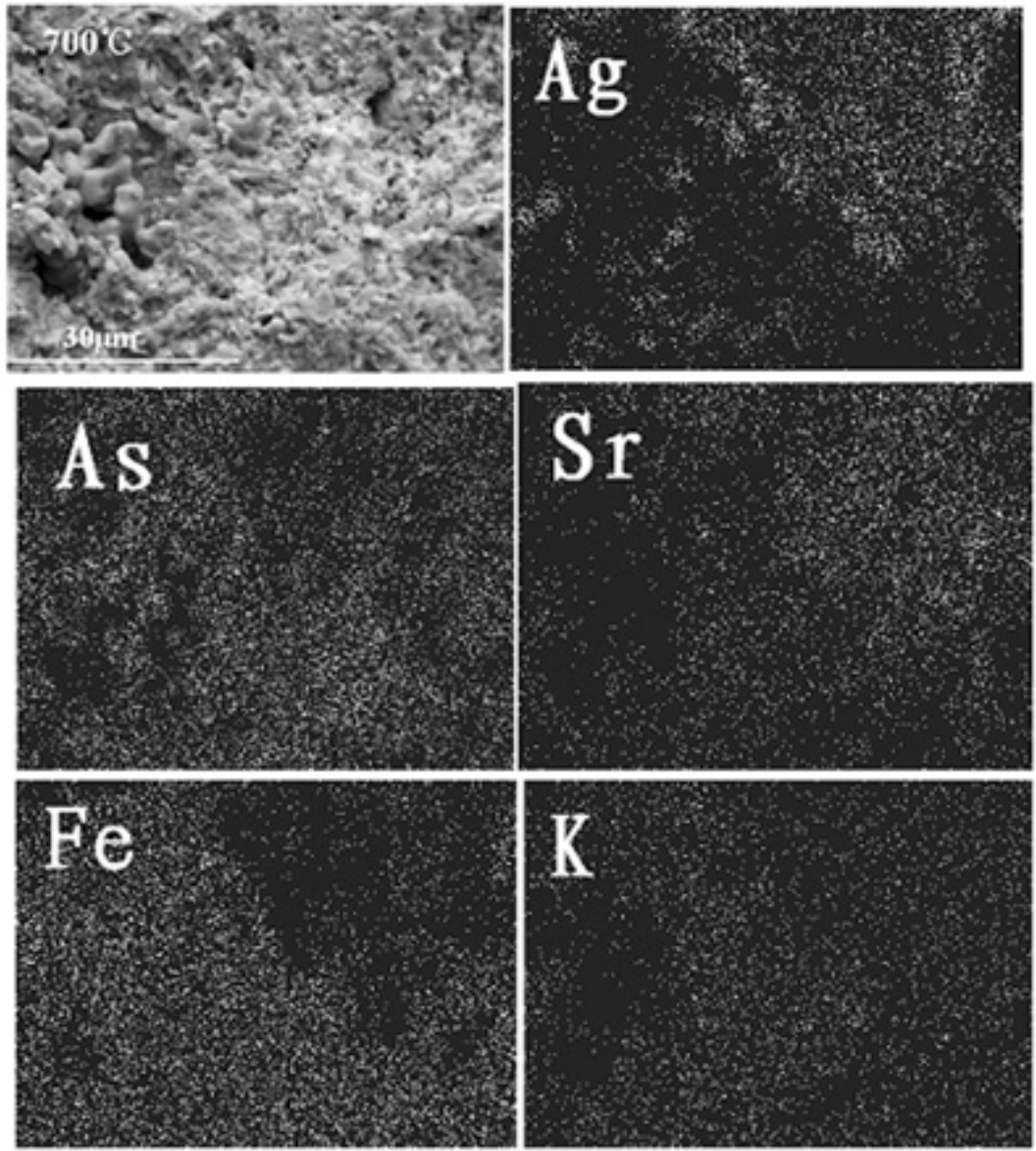

Fig.6 Zhang et al.

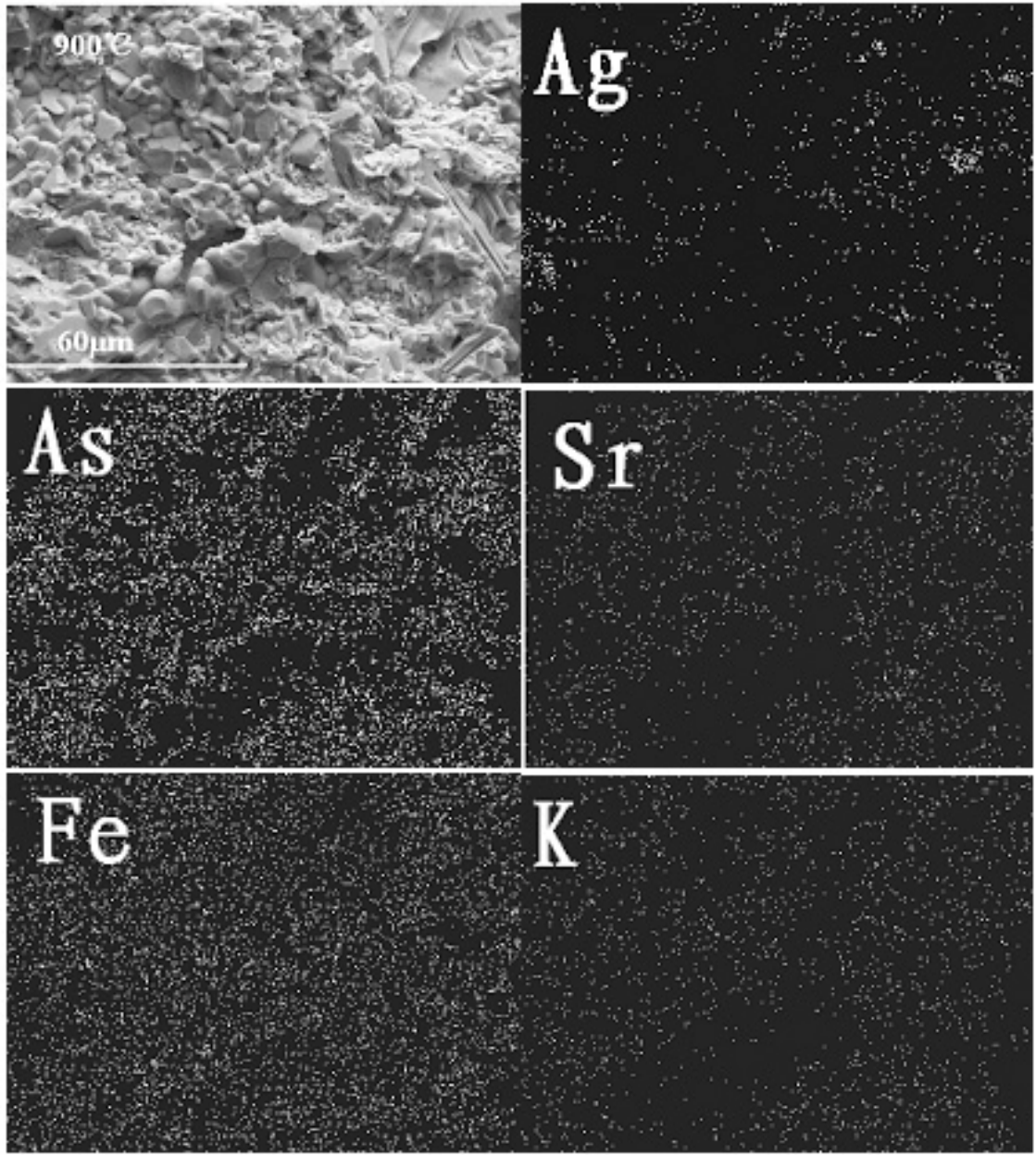

Fig.7 Zhang et al.